# Optically Tunable Threshold Switching and Thermally Activated Transport in Planar Ag/$MAPbI_3$ Thin Single-Crystal Devices

Ofelia Durante[1, §], Valeria Demontis [2,§,*], Sebastiano De Stefano[1], Selene Matta[2], Adolfo Mazzotti[1], Daniela Marongiu[2], Emanuele Meloni[2], Elisa Pili[2], Fang Liu[2], Nicola Sestu[2], Angelica Simbula[2], Mauro Carta[3], Michele Saba[2], Andrea Mura[2], Massimiliano Di Ventra[4], Giovanni Bongiovanni[2], and Antonio Di Bartolomeo[1,*]

*[1] Department of Physics 'E.R. Caianiello', University of Salerno, Via Giovanni Paolo II 132, Fisciano (SA) 84084, Italy*
*[2] Department of Physics, University of Cagliari, Monserrato (CA) 09042, Italy*

*[3] Department of Mechanical, Chemical and Materials Engineering, University of Cagliari, Cagliari, 09123, Italy*
*[4]Department of Physics, University of California San Diego, La Jolla, CA 92093, USA*

§shared authorship

*corresponding author: vdemontis@dsf.unica.it; adibartolomeo@unisa.it

**Abstract**

Halide perovskites have enabled major advances in optoelectronics, extending well beyond photovoltaics. Their mixed ionic-electronic conduction, once regarded as detrimental to device stability, is increasingly viewed as a functional degree of freedom for memory and neuromorphic-inspired devices, especially when coupled to external stimuli such as light. Specifically, single crystals are attractive models because they suppress grain-boundary effects and microstructural disorder that can mask intrinsic transport and interfacial mechanisms in polycrystalline films. Here, we report the growth of thin methylammonium lead iodide ($MAPbI_3$) single crystals by a space-confined method and their integration into planar two-terminal devices with directly deposited Ag contacts. At room temperature, the devices exhibit ultra-low dark currents ($10^{-13}$-$10^{-12}$ A) and negligible hysteresis in the dark. Under illumination, the current increases due to photogeneration and the I-V characteristics develop a pronounced polarity-dependent hysteresis and a threshold-like transition between two conductance states. Temperature-dependent dark measurements (300–400 K) show thermionically activated, contact-influenced transport and a weakly varying normalized hysteresis metric. Together with the back-to-back Schottky-diode analysis and control devices using more inert contact materials, these results support a transport model in which Ag/perovskite interfaces play a central role and the hysteretic response is influenced by coupled interfacial and ionic processes.

*Keywords: $MAPbI_3$, single crystal, space confined growth, hysteresis, ionic migration*

## 1. Introduction

After reshaping the field of photovoltaics, halide perovskites are now impacting other areas of optoelectronic research, catalysing advances in light emission [1,2], photodetection [3], X-ray detection [4], and are emerging as highly promising platforms for neuromorphic electronics [5,6]. In fact, their intrinsic mixed ionic-electronic transport, long considered a challenge for device performance and stability, is increasingly recognized as a powerful functional asset for emulating synaptic dynamics [5,7]. In particular, the exceptional photoresponse accompanied by pronounced dynamical phenomena, such as current-voltage hysteresis, slow transients and persistent photocurrent [8], related to the coupling between electronic and ionic transport, trap dynamics and interfacial electrostatics, enable conductance modulation, switching behaviors and plasticity, that can be tuned electrically and optically, making these materials very interesting for novel neuromorphic devices such as memristors. Moreover, stochasticity and noise-related effects are increasingly recognized as relevant ingredients in memristive dynamics and can strongly influence hysteresis and state stability, motivating dedicated investigations in a variety of material platforms [9–11]. Most studies have so far focused on thin polycrystalline films, where grain boundaries and defects in the surface and in the bulk dominate charge transport, creating poorly controlled channels for ion migration and obscuring the intrinsic properties of the material. Single-crystals (SC), on the contrary, are particularly powerful platforms to investigate intrinsic-transport and defect-mediated processes because they minimize microstructural disorder and grain-boundary effects that complicate the study of these materials. At the same time, SC based devices still display substantial history-dependent electrical behavior because ionic redistribution and interfacial charging can occur even in the absence of grain boundaries[12]. Despite the progress in recent years, the synthesis of solution-based high quality single-crystal perovskites, especially in the thin film form (thickness below a few micrometers), which is the only suitable form for device integration and scalability, is still not a very mature and controlled process [12,13].

In this work we grow methylammonium lead iodide ($CH_3NH_3PbI_3$-$MAPbI_3$) single-crystals using a space-confined inverse temperature approach and integrate them into planar, optically accessible, two terminal devices using an easy and low-cost fabrication approach based on the direct deposition of conductive silver electrodes. We show that these devices under illumination show strong photoconductivity, pronounced polarity-dependent hysteresis, and a threshold-like transition between two conductance states that can be tuned by the incident optical power. To rationalize this behaviour, we analyse the transport within a back-to-back Schottky barrier framework, which captures the contact-limited character of the devices and the dominant role of the Ag/$MAPbI_3$. In addition to the light-induced effects, temperature-dependent measurements in the 300–400 K range confirm thermally activated transport and contact-modulated transport. These results highlight the importance of the study of light stimuli in tuning memristive-like behaviours in halide perovskites and show that Ag electrodes can act as active electrochemical interfaces enabling light-assisted barrier modulation and threshold-like switching. This view is consistent with recent reports emphasizing the role of Ag-based electrode engineering in halide-perovskite devices [14,15].

## 2. Results and discussion

### 2.1. Single-crystal synthesis, morphological, structural and optical characterization.

$MAPbI_3$ is a three-dimensional (3D) organic-inorganic hybrid perovskite material composed of a network of corner-sharing halide octahedra with a metal cation at the centre and organic cations occupying the spaces between the octahedra [16,17]. Their characteristic crystal structure is described by $ABX_3$, where A is a small-sized monovalent organic or inorganic cation, B is a divalent metal cation ($Pb^{2+}$ or $Sn^{2+}$), X is a halide anion. SCs-$MAPbI_3$ were grown by utilizing the *space-confined* method as shown in **Figure 1a**. A precursor solution of $MAPbI_3$ 1.3 M in γ-butyrolactone (GBL) was prepared by dissolving lead iodide ($PbI_2$) (from Sigma Aldrich) and methylammonium iodide (MAI) (from Greatcellsolar) in a 1:1 molar ratio and maintaining the mixture in a hot plate at 65°C under continuous stirring for 2 h.

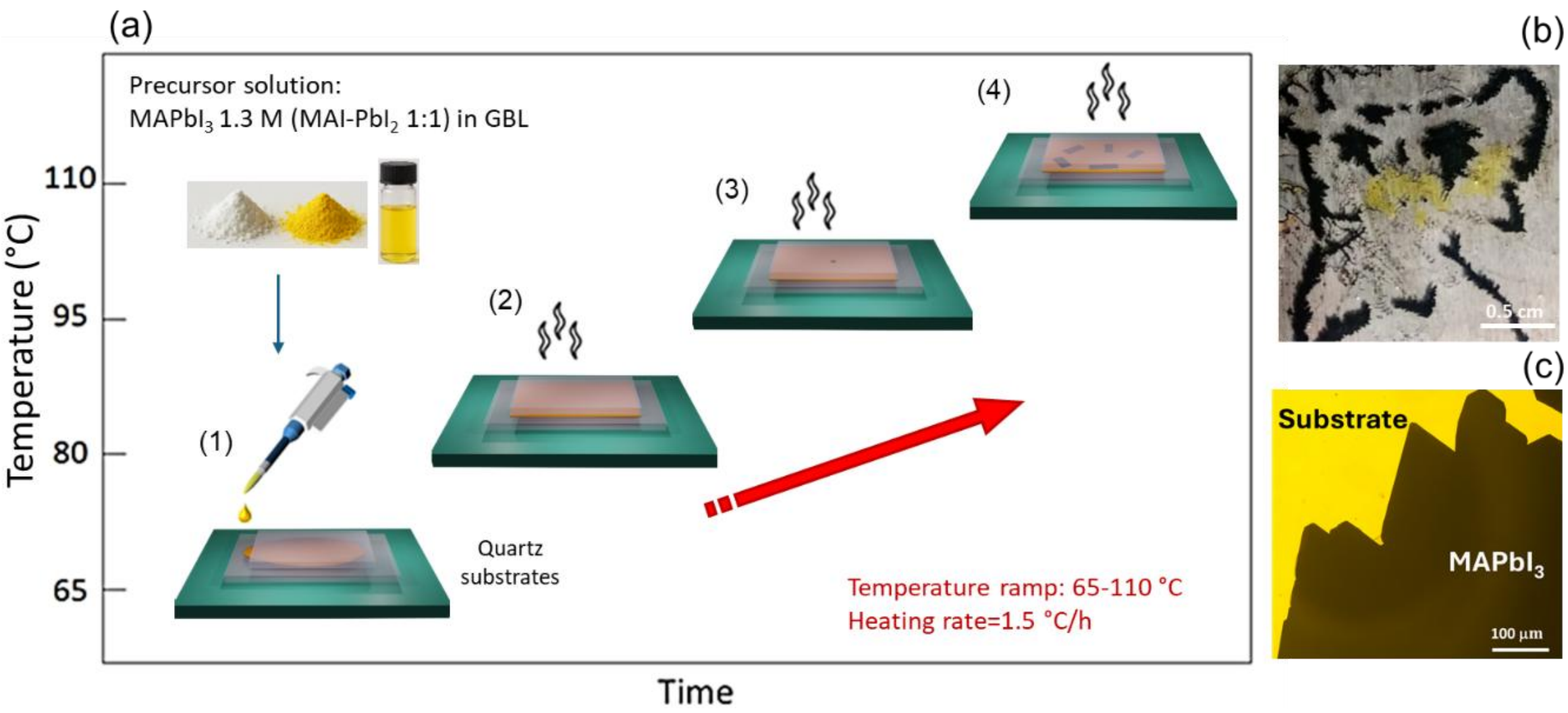

**Figure 1.** (a) Schematic representation of the space-confined method. After substrates preheating for 10 min at 65°C in a programmable oven, 10 µL of 1.3M MaPbI3 precursor solution in GBL (MAI/PbI2 1:1) are injected in the space between two quartz substrates (1); the temperature is then raised from 65°C to 110°C at a rate of 1.5 °C/h. As heating proceeds, the solubility of the precursor decreases (2) due to reverse solubility and 3D perovskite starts nucleation (3). Then, solvent evaporation leads to a higher concentration of the remaining solution, sustaining supersaturation and promoting crystal growth (4). Different nucleation sites result in several single crystals with random distribution in the substrate area and random size. (b) Photograph of a grown sample after full solvent evaporation, obtained keeping the sample in dynamic vacuum for about two weeks. The substrates are then be separated using a blade. (c) Bright-field transmission optical image of one of the grown crystals.

Prior to growth, quartz substrates measuring 1.5 cm x 1.5 cm and 2.5 x 2.5 cm were cleaned via sonication in acetone, followed by rinsing in isopropyl alcohol. The substrates were then stacked on top of one another and thermalised in the oven at 65 °C for 10 minutes to avoid thermal shock. 10 µL of the precursor solution was then infiltrated in the space between the substrates by capillarity, by depositing droplets at the border of the upper substrate. The temperature was then gradually increased at a constant rate of 1.5°C/h from 65°C to 110 °C to promote the material's crystallization exploiting the inverse temperature crystallization of this material in GBL [18]. The samples were then kept at a constant temperature of 110 °C for 24h to further promote the increase the crystals size. At the end of the process, to facilitate the solvent evaporation and yield dry films, the samples were kept under dynamic vacuum conditions in a vacuum chamber connected to a scroll pump [19]. Before use, the two substrates are separated (**Figure 1b**), resulting in crystals with dimensions reaching the millimetre scale. **Figure 1c** displays a bright-field transmission micrograph of a representative $MAPbI_3$ crystal. The crystal has a characteristic dark contrast due to strong absorption in the visible range. The sharp contours and plate-like morphology, indicate a well-formed crystalline domain. The room temperature X-ray diffraction (XRD) pattern, **Figure 2a**, shows the comparison between $MAPbI_3$ single crystal diffraction patter (red) with its powder (gray). The single crystal configuration is confirmed by the presence of intense diffraction peaks at 14° and 28°, which can be assigned to the tetragonal phase of $MAPbI_3$, as highlighted in the inset. These reflections can be indexed as the (110), (220) … (hk0) from parallel planes, indicating a strong preferred orientation of the crystals, with dominant crystal growth along (hk0) direction, with the c-axes parallel to the substrate. In contrast, the powder sample exhibits several diffraction peaks associated with multiple crystallographic orientations.[16,20]. **Figure 2b** shows the room-temperature photoluminescence (PL) spectrum under 630 nm excitation. The spectrum is characterized by a main peak cantered at 773 nm corresponding to an energy of ~ 1.6 eV, i.e., close to band-edge emission of $MAPbI_3$ [21].

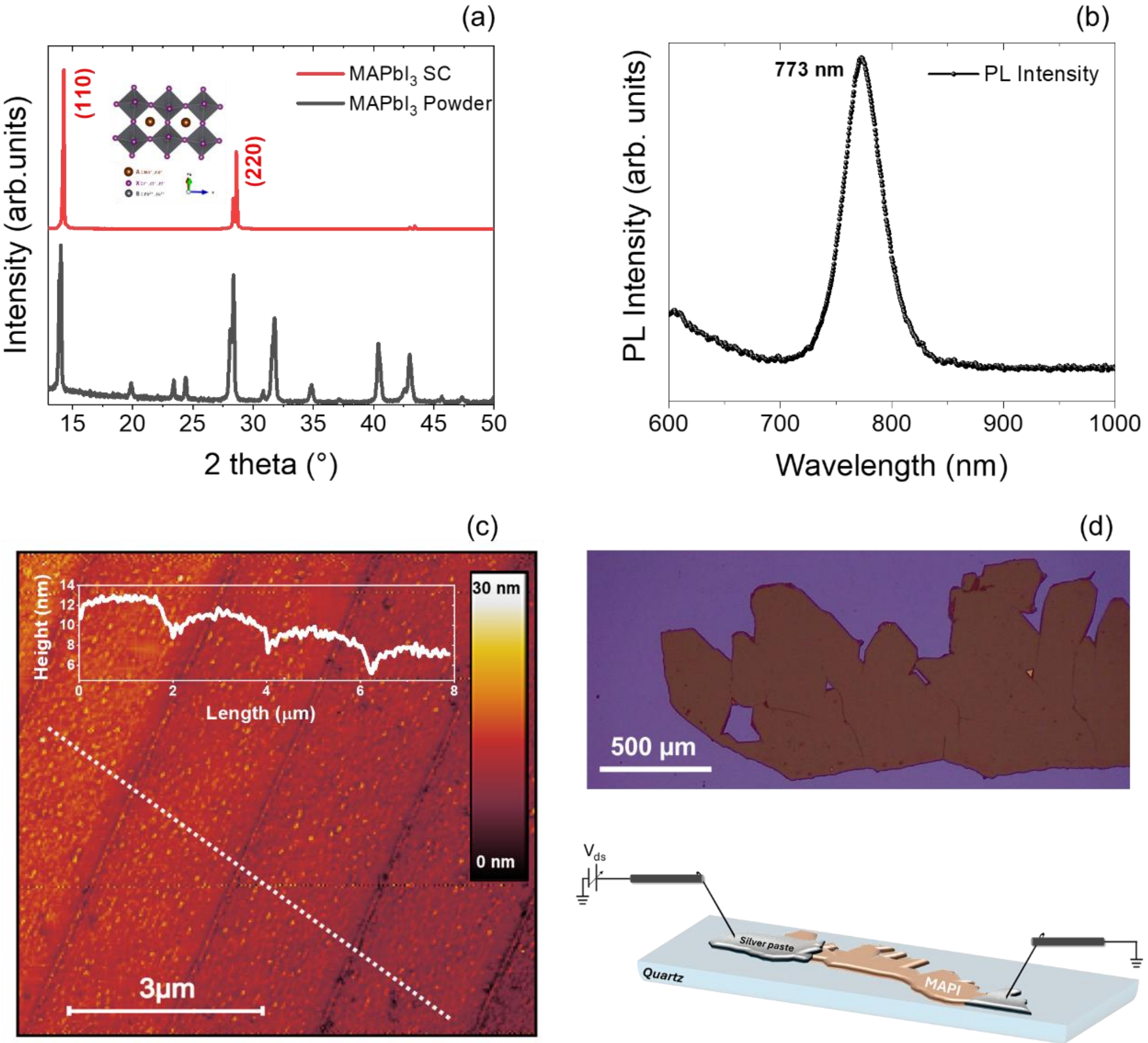

**Figure 2.** (a) Comparison between the XRD patterns of $MAPbI_3$ single crystal (red) and corresponding powder sample (gray) (inset: crystal structure of $MAPbI_3$); (b) Photoluminescence (PL) under 630 nm excitation of $MAPbI_3$ single crystals. (c) 8x8 $\mu m^2$ AFM-scan area and profile (inset) of $MAPbI_3$ single crystals (bottom). Optical image of the single crystal (d) after the growth (top) and schematics of the device with silver electrodes. (bottom).

**Figure 2c** shows the Atomic Force Microscopy (AFM) topography, on a scan area of $8 \times 8\ \mu m^2$, of the $MAPbI_3$ single-crystal surface. The image reveals a relatively smooth surface with the presence of several well-defined terraces [22], Inset of **Figure 2c**, which are characterized by step heights of ~3.7 nm. Finally, **Figure 2d (top)** shows the optical images of the SC-$MAPbI_3$ used for device fabrication. Specifically, planar devices were fabricated by directly depositing a droplet of a custom-made silver paste at each crystal edge. The conductive paste was prepared from a commercial DuPont 4929N silver paste, whose proprietary formulation mainly contains Ag particles, organic binder (methyl methacrylate) and solvent. To avoid direct exposure of $MAPbI_3$ to the original paste solvent, the solvent was fully evaporated and the remaining solid fraction was re-dispersed in chlorobenzene. The paste was stirred for 2 h, deposited at the crystal edges, and thermally cured for 15 min at 100 °C. The resulting Ag contacts have an estimated thickness of ~100 μm and a geometrical contact area of ~0.06 $mm^2$.

The final device is thus characterized by a channel width of ~400 μm and an electrode gap (channel length) of ~1.5 mm (see sketch in **Figure 2d-bottom**).

## 2.2 Photoresponse at room temperature

**Figure 3a** displays the current-voltage (I–V) characteristics measured in dark at room temperature and at pressure of 1.8 mbar. The device shows an ultra-low current ($10^{-13}$ - $10^{-12}$ A) in the bias range investigated, due to the low intrinsic carrier concentration in single-crystals [23]. Moreover, in dark conditions, the forward/backward I-V curves are almost overlapping, indicating quasi-stationary transport and a limited contribution of slow processes, such as ionic migration and redistribution and charge trapping during the sweep; the absence of grain boundaries in single crystals likely suppresses one major pathway for ion-assisted hysteresis observed in polycrystalline films [24]. The quasi-linear dark I–V behaviour (inset) indicates that, within the investigated bias range, the response is dominated by the highly resistive $MAPbI_3$ channel, which masks the rectifying contribution of the contact barriers. From the linear fit of the dark I–V curve (red dashed line in the inset), we extracted a channel resistance of 9.4 x $10^{11}$ Ω.

Upon illumination with a supercontinuum laser (450–2400 nm), the device exhibits a strong photocurrent, with $I_{on}/I_{off}$ current ratio exceeding $10^4$. As shown in **Figure 3b,** the current increases systematically with incident laser power, $P_{inc}$, calculated as $P_{inc} = (P/A_{spot})\cdot A$, where $P/A_{spot}$ is the laser irradiance (11 W $cm^{-2}$ at the maximum power) and $A$ is the active device area (0.003 $cm^2$), indicating that the response is dominated by photoconductive effects (the linear behaviour of the photocurrent as function of the incident power is confirmed in the inset of **Figure 3b).** Moreover, **Figure 3b** shows that, under illumination, a polarity-dependent asymmetry emerges in the I-V curves together with a pronounced hysteresis at negative bias (or at positive bias when the contacts are inverted as reported in **Figure S1** of Supporting Information).

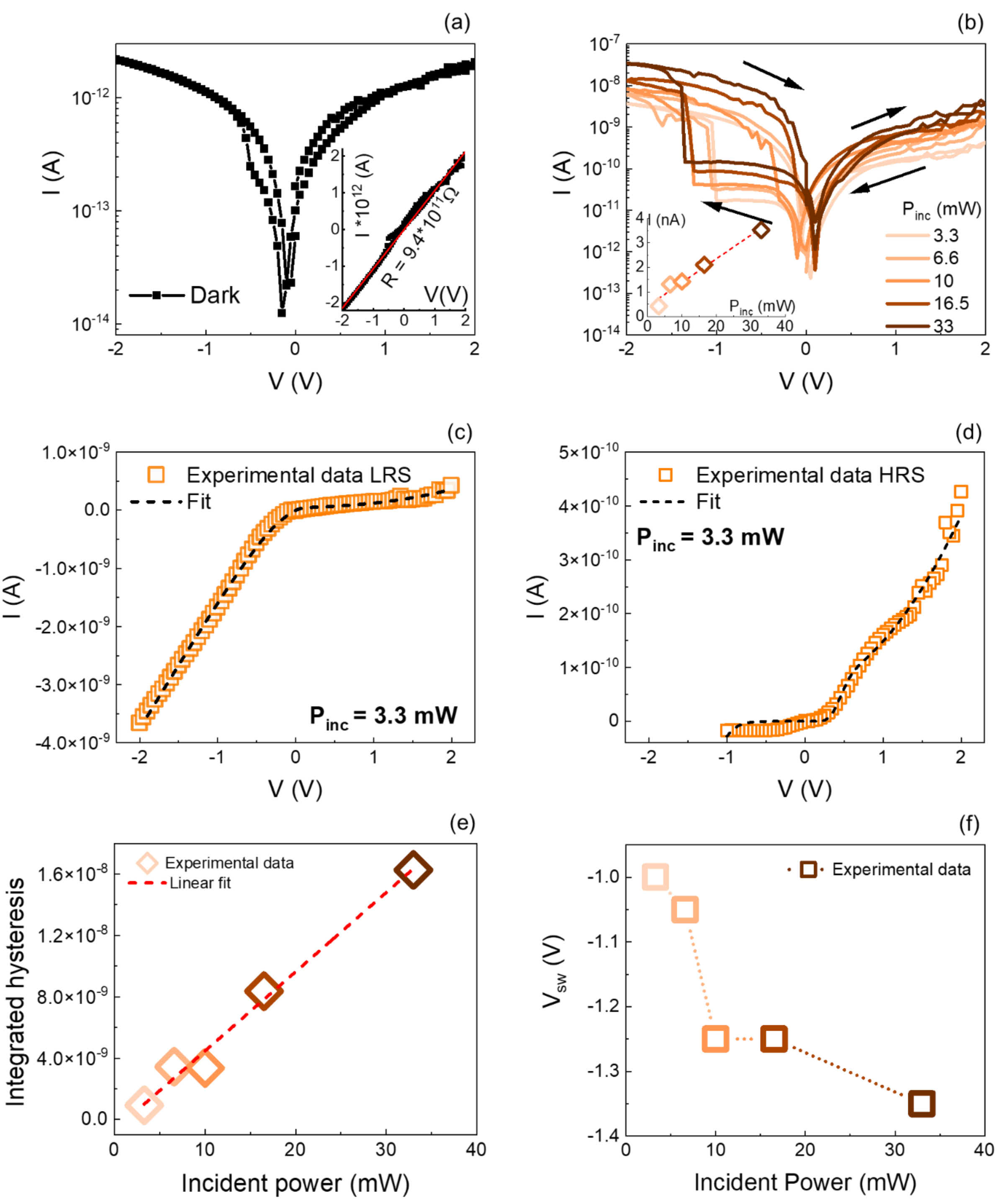

**Figure 3.** (a) I–V curves on a semi-logarithmic scale measured in darkness (dark). Inset: I-V on linear scale. (b) I–V curves under supercontinuum laser illumination at different incident powers, where the dark arrow indicates two current levels ($I_{low}$ and $I_{high}$).Inset: Photocurrent extracted at V = +2 V vs incident power with a linear fit (red). I–V curve (orange) compared with the fit (black) for (c) low-resistance state (LRS) and (d) high-resistance state (HRS) obtained using a back-to-back Schottky diode model including series resistance $R_s$. (e) Integrated hysteresis vs incident power. The red dotted line indicates a linear fit. (f) $V_{sw}$ vs incident power. The dotted line acts as a guide for the eyes.

These phenomena are consistent with the mixed ionic-electronic nature of halide perovskites, in which light and bias can activate history-dependent responses through coupled ionic polarization and electronic conduction, resulting in delayed current dynamics [25]. Specifically, the threshold switching-like behaviour behavior appears to be enabled by the presence of Ag contacts, which are known to induce light-assisted reactions at the metal/perovskite interface, leading to the formation of mobile Ag+ ions species. These species modify the injection barrier, and their dynamics combine with photogeneration and light activated intrinsic ions dynamics (usually charged iodine defects, which have the lowest activation energies) [26]. As displayed in **Figure S2** of Supporting Information, in the investigated range, the hysteresis is negligible in illuminated single-crystal $MAPbI_3$ devices with graphite- and gold-based electrodes, which are known to be much more inert materials when in contact with perovskites [27,28]. Indeed, the study of controlled silver diffusion in perovskite layers, is nowadays regarded as a powerful tool to enhance device performances, enable novel functionalities and develop new passivation strategies [14,15,29].

The strong hysteresis in the curves reported in **Figure 3b** suggests sharp transitions between two resistance states at specific switching voltages, $V_{sw}$, with the separation between them increasing as the light power increases. The hysteresis is present only in branch of the I-V curve while the other remains substantially unaffected, suggesting that the switching behaviour is dominated by an asymmetric interface process at the contacts rather than a spatially homogenous bulk response. Moreover, the formation of a high conductive filamentary-type channel in the material, often invoked in related systems, appears unlikely in the present case. On the one hand, the channel length is on the order of several hundreds of micrometres, making the establishment of conductive filamentary pathways in the channel, such as Ag ions percolation pathways, very improbable. On the other hand, the measured current levels remain several orders of magnitude lower than those typically reported for filamentary switching phenomena [6]. Instead, the observed behaviour is more likely related to the presence of two Schottky diodes at the Ag/$MAPbI_3$ interfaces[30], not equally rectifying as commonly observed in small scale semiconductor devices, and to a light and voltage assisted abrupt change in the conductivity of one of the contacts, driving the transition from a high resistance state to a low resistance state.  To support this interpretation, we verified that both forward and reverse-sweep I-V curves under illumination can be very well reproduced by modelling the system as a two-terminal device with rectifying contacts, described by an equivalent circuit consisting of two back-to-back diodes separated by a series resistance[31]:

$$I = \frac{2I_{s1}I_{s2}\,sinh\left(\frac{qV}{2kT}\right)}{I_{s1}e^{\frac{qV}{2kT}} + I_{s2}e^{-\frac{qV}{2kT}}}$$

where $q$ is the elementary charge, $k$ is the Boltzmann constant and $T$ is the temperature. The quantities $I_{s1}$ and $I_{s2}$ are the reverse saturation currents of the two junctions, written as:

$$I_{s1,s2} = S_{1,2}\, A^{*}T^{2}\, exp(-\frac{\Phi_{B01,2}}{kT})$$

with $A*$ the Richardson constant, $S_{1,2}$ the effective areas of the junctions, and $\Phi_{B01,B02}$ the ideal Schottky barriers at zero bias. The effective Schottky barrier, which takes into account deviations from ideality due to defects and image-force lowering that makes the Schottky barrier height dependent on the applied external voltage, can then be written as:

$$\Phi_{B1,B2} = \Phi_{B01,B02} \pm eV_{1,2}\left(1 - \frac{1}{n_{1,2}}\right)$$

where $n_1$ and $n_2$ are the ideality factors of the two diodes, and $V_{1,2}$ denotes the voltage drop across each junction and the sign depends on the junction polarity under bias. To include the voltage drop along the $MAPbI_3$ channel, we introduce a series resistance $R_s$ and define the effective voltage as:

$$V = V_{app} - I\,R_S$$

where $V_{app}$ is the applied external voltage (similar simulation is done on the dark I-V shown in **Figure S3**). **Figure 3(c-d)** show the comparison between the experimental and fitted curves for one of the I-V characteristics reported in **Figure 3b** (under $P_{inc}$ = 3.3 mW), confirming that the back-to-back Schottky diode model provides a good explanation for the device behavior (the main fitting parameters are reported in Table S6 of Supporting Information for both the high-resistance and low-resistance states).

We highlight that, in dark, the very low conductivity of $MAPbI_3$ single crystals leads to an approximately linear response because the voltage drop along the channel dominates over the drop at the contacts. Under illumination, the channel becomes much more conductive and the rectifying role of contacts becomes relevant. Likely, under illumination, the voltage drop at one of the contacts becomes high enough to enable silver-ion migration at the interface, causing the rapid modification of the barrier and making the junction quasi-ohmic, as observed in the negative branch of the IV curve in **Figure 3c**. This is also supported by Energy Dispersive X-ray Spectroscopy (EDS) analysis reported in Supporting Information. In the pristine device (Figure S4 a,b), the Ag signal is confined to the electrode region and rapidly drops to the background level in the perovskite. After biasing and transition to the low-resistance state (Figure S4 c,d), a weak residual Ag signal is detected in the $MAPbI_3$ region adjacent to the contact, suggesting bias-induced Ag diffusion near the $MAPbI_3$ interface. When the sweep is reversed toward positive bias, the conductive interfacial region is expected to dissolve and the barrier height to recover its original value. However, this reverse transition is not directly resolved in the measured I-V curve, because in that bias range the current is limited by the other Schottky junction, which is reverse-biased and masks the recovery of the switched contact. Moreover, the switching is not perfectly bipolar, most likely because the two contacts are not strictly symmetric due to fabrication-related differences.

To quantify the evolution of the hysteresis with illumination, and thus the amplitude of the memory window in the memristive device, we introduce an integrated-hysteresis metric defined as the area between the forward and backward sweeps:

$$A_{\mathrm{hyst}} = \int \mid I_{\mathrm{fw}}(V) - I_{\mathrm{bw}}(V) \mid dV.$$

As shown in **Figure 3e**, $A_{\mathrm{hyst}}$ increases monotonically with illumination, indicating a progressive opening of the I–V loop and an enhanced sweep-to-sweep separation as photoexcitation becomes stronger. This behaviour shows that illumination amplifies the history-dependent component of transport, i.e., the difference between forward and backward branches. Within a contact-limited back-to-back Schottky framework, this trend is naturally explained by the fact that the reverse current depends exponentially on the effective barrier height; under illumination, the channel becomes more conductive and a larger fraction of the applied bias drops at the Ag/$MAPbI_3$ interfaces, making interfacial barrier modulation (via interfacial charging and thermally/optically activated ionic redistribution, such as $Ag^+$ ions dynamics at the interface) more effective during the sweep. As a result, small illumination-induced variations of effective barrier height can produce large changes in the reverse-limited current and therefore a marked increase of the integrated hysteresis area with optical power. We further show the switching voltage, $V_{sw}$, defined as the bias at which the current exhibits the steepest transition between the two states. $V_{sw}$ shifts progressively toward more negative values as illumination increases (**Figure 3f**). This trend is consistent with a contact-limited interfacial scenario in which illumination enhances carrier injection and interfacial charging/ionic polarization, thereby modifying the effective Schottky barriers and the condition for triggering the transition. At the same time, we note that a purely kinetic/time-dependent contribution cannot be excluded: if the transition requires a finite activation time (e.g., for ion/Ag redistribution at the interface), higher currents under stronger illumination may lead to a faster evolution of the internal state and to an apparent shift of the observed $V_{sw}$ during a continuous voltage sweep. All I-V curves reported in this work were acquired using the same voltage-sweep protocol, corresponding to an effective scan rate 230 mV/s. Therefore, the extracted $V_{sw}$ values should be regarded as effective thresholds under fixed sweep conditions rather than scan-rate-independent quantities.

It is worth noticing that, as largely acknowledged in the field of halide perovskite memristors, the control and reproducibility of switching mechanism is still an open challenge. In our experiments, out of five measured devices we observed the reported behavior in three samples fabricated with the same process, also using different $MAPbI_3$ synthesis batches. In one sample we observed switching at both positive and negative voltages, with a stochastic value of the threshold voltage and poor light control, and another one didn't display any switching.

We also note that the contact asymmetry underlying the observed mechanism can be systematically introduced through appropriate contact engineering. For instance, using contacts made of different materials or adopting a geometry in which the metal lies beneath the perovskite at one contact and above it at the other can introduce a systematic and controllable asymmetry.

### 2.3 Current versus temperature

**Figure 4a** shows the semilogarithmic dark I-V curves acquired between 300 and 400 K at pressure of 1.8 mbar. We note that $MAPbI_3$ undergoes the tetragonal-to-cubic structural transition in the same temperature window [32–34]. Diffraction studies further indicate that the transition is first-order and can display a sizeable cubic/tetragonal phase-coexistence region (order of 10–30 K) around $T_c$ [34]. As already mentioned, the I-V characteristics are compatible with the behaviour of the thermionic current in back-to-back Schottky diode configurations. Accordingly, at fixed bias, the dark current increases with temperature, as expected for predominantly contact-limited, thermally activated (thermionic) injection over interfacial barriers. In addition, the separation between forward and backward sweeps, negligible at room temperature, becomes more pronounced at higher temperature, consistent with temperature-accelerated internal-state dynamics that modulate the effective contact barriers.

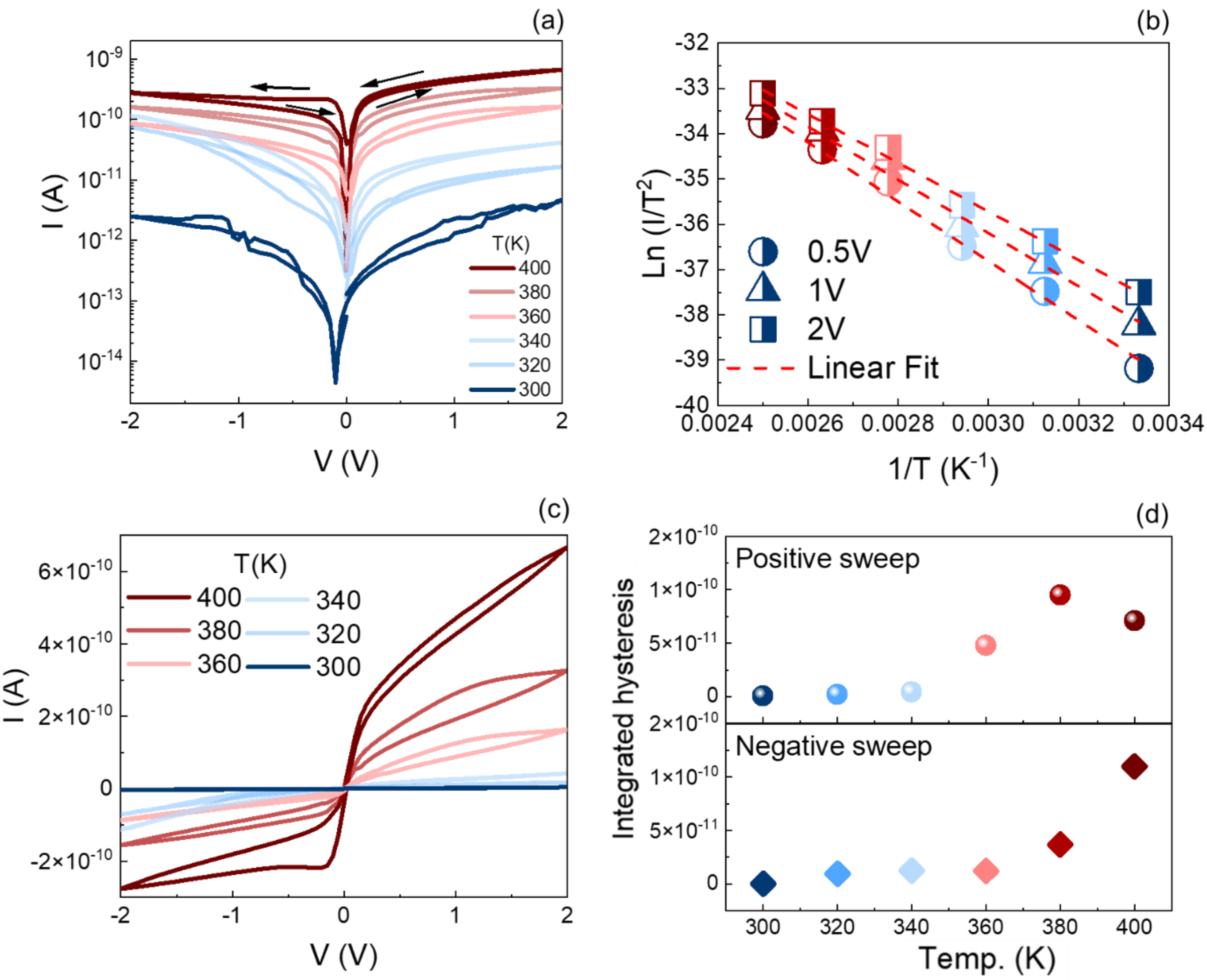

**Figure 4.** (a) I-V curves on a semi-logarithmic scale at different temperatures. (b) Richardson plot, ln(I/T$^2$) vs $1/T$ for different voltage bias (0.5, 1, 2 V). (c) I-V curves on a linear scale at different temperatures. (d) Integrated Hysteresis for positive (top) and negative (bottom) sweep as a function of temperature.

Despite the increase in hysteresis amplitude, the sharp transition to a higher-conductance state observed under illumination is not observed in the dark, even at higher temperatures. This is consistent with a contact-limited scenario in which, in the dark, the $MAPbI_3$ channel remains poorly conductive so that a significant fraction of the applied bias drops across the bulk rather than being concentrated at the metal/semiconductor interfaces; as a result, the effective field at the contacts may remain insufficient to trigger the interfacial/ionic rearrangements required for the abrupt transition.

To quantify the temperature dependence, the current was extracted from positive-bias branch at $V = +0.5, +1$ and $+2\ V$ and analysed within a contact-limited thermionic-emission framework. In the presence of Schottky barriers at the metal/$MAPbI_3$ interfaces, the current is expected to follow a Richardson-type law [35,36]

$$I(T) \propto T^2 \exp\left[-\frac{q\Phi_B}{k_B T}\right],$$

so that a linear dependence is obtained by plotting $ln(I/T^2)$ versus $1/T$. As shown in **Figure 4b**, the experimental data are well described by this linearization, allowing us to extract an effective Schottky barrier height $\Phi_B$. From the slopes, we obtain $\Phi_B = 0.56\ \pm 0.03\ eV$ at 0.5 V, $0.51\ \pm 0.03\ eV$ at 1 V, and $0.46\ \pm 0.02\ eV$ at 2 V. A corresponding Richardson analysis for the negative-bias branch, reported in the Supporting Information (**Figure S5**), confirms the same contact-limited thermally activated transport picture. The systematic decrease of $\Phi_B$ with increasing bias indicates a field-assisted and contact-influenced conduction regime rather than purely intrinsic bulk transport. In halide perovskites, mixed ionic-electronic conduction and interfacial polarization can dynamically modulate band bending and effective injection and extraction barriers, producing history-dependent currents. In this context, the extracted $\Phi_B$ values ($\sim 0.46 - 0.56$ eV) are compatible with interfacial and defect-mediated scenarios in which iodine-related point defects and interfacial trapping affect the effective barrier at the $MAPbI_3$ surface and its release kinetics is thermally activated, as proposed by Caddeo *et al.* [37]. Finally, **Figure 4c** shows the dark I–V curves on a linear scale, highlighting the sweep-to-sweep separation of the current branches as temperature increases. To quantify this effect, we evaluate the hysteresis area as the integral of the difference between the forward and backward sweeps for the positive and negative bias, defined as:

$$A_{\text{hyst}}^{-} = \int_{-V_{\max}}^{0} | I_{\text{fw}}(V) - I_{\text{bw}}(V) | \; dV$$

and

$$A_{\text{hyst}}^{+} = \int_{0}^{V_{\max}} | I_{\text{fw}}(V) - I_{\text{bw}}(V) | \; dV.$$

**Figure 4d** reports the integrated hysteresis areas extracted separately for the positive (top) and negative (bottom) bias halves of the sweep. $A_{\text{hyst}}^{+}$ is already finite at 300–340 K and increases markedly above ~360 K, whereas $A_{\text{hyst}}^{-}$ remains negligible up to ~340 K and becomes significant only at higher temperature. This polarity-dependent activation is consistent with a back-to-back Schottky, contact-limited picture in which thermally accelerated interfacial/ionic dynamics modulate the effective barriers asymmetrically at the two Ag/$MAPbI_3$ interfaces, with one interface dominating at low temperature and both contributing at higher temperature. Moreover, this temperature-driven evolution is qualitatively consistent with general memristive frameworks in which diffusion-controlled internal dynamics can enhance hysteresis [11].

## 3. Conclusions

In this work, thin $MAPbI_3$ single-crystals, grown by using space-confined method, were integrated into planar two-terminal devices with Ag contacts. In the dark at room temperature, the devices exhibit ultra-low currents ($10^{-13}$ - $10^{-12}$ A) and negligible sweep-to-sweep separation, consistent with suppressed microstructural contributions in single crystals. A pronounced polarity dependent, threshold-like transition between two conductance states emerges in the illuminated I-V curve. The switching amplitude increases with incident optical power and the switching voltage shifts progressively toward more negative bias, demonstrating that the resistive response can be tuned by light. The asymmetry of the illuminated hysteresis, its dependence on the electrode material, and the agreement with a back-to-back Schottky description support the view that switching is governed primarily by interface-mediated barrier modulation rather than by a spatially homogeneous bulk process. Temperature-dependent dark I–V measurements further show thermionically activated, contact-influenced transport between 300 and 400 K; while the absolute current increases strongly with temperature, a normalized integrated-hysteresis metric varies only weakly, indicating that history-dependent internal-state contributions remain a secondary effect in the dark conditions. Finally, planar Ag/$MAPbI_3$ thin single-crystal devices provide a useful testbed to study light- and temperature-controlled resistive phenomena in halide perovskites and to clarify the active role of metal/perovskite interfaces in memristive behaviors.

## Experimental Section

The optical microscope used for bright-field analysis was a Nikon Eclipse TE-2000U. The image was captured under white-light in transmission mode using a CS126CU Kiralux 12.3 MP Color CMOS Camera with USB 3.0 interface.

The perovskite X-Ray Diffraction pattern was measured with a Bruker D8 ADVANCE diffractometer employing a Cu Kα radiation (40 kV/40 mA, $\lambda$= 1.5418 Å) and a Goebel mirror with a UBC 0.3 mm collimator. The sample was placed on a motorized XYZ stage (UMC stage), and data were collected with the LYNXEYE XE-T detector in θ–2θ mode at 0.025–0.03° step size and 0.5 s per step.

The surface morphology and associated terraces profile of the three-dimensional $MAPbI_3$ single crystal were characterized using a Horiba AINST-NT Atomic Force Microscope in non-contact mode, employing an HQ:NSC19/Al BS cantilever with a nominal force constant of 0.5 N/m. All acquired data were subsequently processed and analysed with Gwyddion software.

Scanning electron micrograph and EDS analysis were performed using a Coxem EM-40 scanning electron microscope, equipped with Bruker Quanta ED-XS detector.

The steady-state photoluminescence spectrum of the crystals was acquired by exciting the sample with a Phaos laser combined with an Orpheus system at 630 nm and 50 kHz repetition rate, and acquiring the signal in reflection configuration using a SPM-002 fiber spectrometer and a RG665 filter.

Electrical measurements were performed using a Janis probe station (Janis ST-500 probe station), equipped with tungsten probes, with optical microscope, and with temperature control and a pressure range from 1 to $10^3$ mbar. Keithley 4200 SCS (semiconductor characterization system), working as a source-measurement unit with current sensitivity better than 1 pA, was used for electrical characterization. The photoresponse was tested using a SuperK COMPACT supercontinuum white-light lasers by NKT Photonics with a spectrum spanning 450-2400 nm and an optical power of up to 110 mW (nominal value).

**Acknowledgments**

V.D. acknowledges the support of the Project NEST++ ′Network 4 Energy Sustainable Transition—NEST′, Project code QIIR112_00040, funded by the Italian Ministero dell'Università e della Ricerca (MUR). V.D., A.S and M.S. acknowledge Ecosystem of Innovation for Next Generation Sardinia, Spoke 7-Project code ECS00000038 (Low carbon technologies for efficient energy system). D.M. acknowledges Fondazione di Sardegna through the project: “Thin Single-Crystal Halide Perovskites for Stable Neuromorphic devices”, F83C2600035000. A.D.B. and O.D. acknowledge the financial support from the University of Salerno, with grants ORSA223384 and ORSA235199. All authors acknowledge CeSAR-Centro Servizi di Ateneo per la Ricerca-at the Università degli Studi di Cagliari (Dr. E. Podda for technical assistance on XRD).

# Supporting Information:

# Optically Tunable Threshold Switching and Thermally Activated Transport in Planar Ag/$MAPbI_3$ Thin Single-Crystal Devices

Ofelia Durante[1, §], Valeria Demontis [2,§,*], Sebastiano De Stefano[1], Selene Matta[2], Adolfo Mazzotti[1], Daniela Marongiu[2], Emanuele Meloni[2], Elisa Pili[2], Fang Liu[2], Nicola Sestu[2], Angelica Simbula[2], Mauro Carta[3], Michele Saba[2], Andrea Mura[2], Massimiliano Di Ventra[4], Giovanni Bongiovanni[2], and Antonio Di Bartolomeo[1,*]

*[1] Department of Physics 'E.R. Caianiello', University of Salerno, Via Giovanni Paolo II 132, Fisciano (SA) 84084, Italy*
*[2] Department of Physics, University of Cagliari, Monserrato (CA) 09042, Italy*

*[3] Department of Mechanical, Chemical and Materials Engineering, University of Cagliari, Cagliari, 09123, Italy*
*[4]Department of Physics, University of California San Diego, La Jolla, CA 92093, USA*

§shared authorship

*corresponding author: vdemontis@dsf.unica.it; adibartolomeo@unisa.it

**Index**

## S.1 Contact polarity inversion

To verify the polarity-dependent nature of the effect and exclude trivial instrumentation artefacts, we repeated the I–V measurement under the same LED illumination conditions after reversing the probe connections (1→2 vs 2→1). The I–V curve is mirror-inverted (**Figure S1**), indicating that the observed asymmetry is linked to the intrinsic device polarity, i.e., to the two Ag/ $MAPbI_3$ interfaces and their non-equivalent effective barriers under bias, rather than to the external measurement setup. Moreover, measurements were acquired consecutively under the same illumination settings.

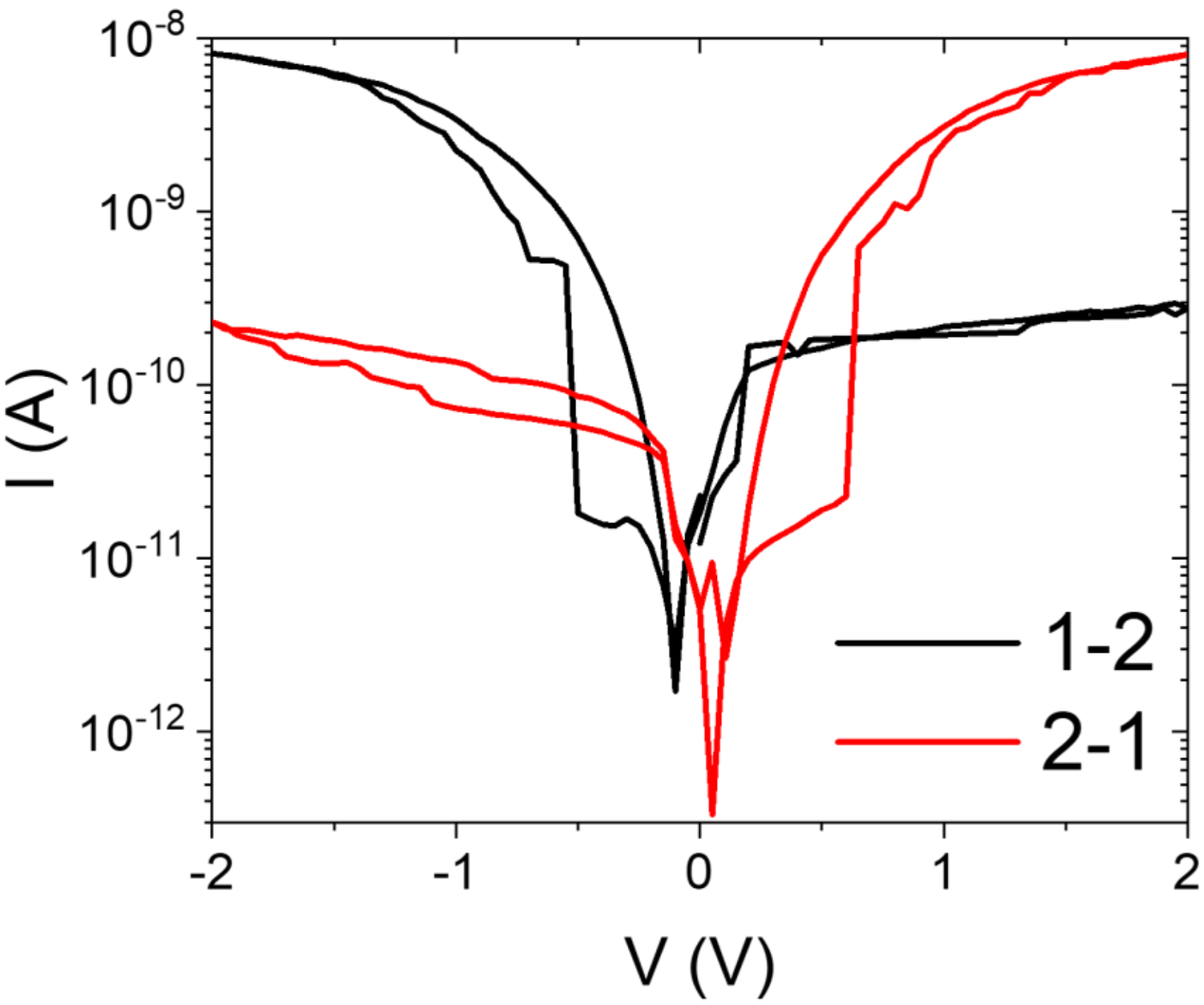

**Figure S1:** I–V characteristics of an Ag/$MAPbI_3$ device measured under identical LED illumination conditions at T=300 K and under pressure of 1.8 mbar with two opposite probe configurations: contact 1→2 (black) and contact 2→1 (red).

## S.2 Role of metal contact

To evaluate the role of the contact material on the transport response of $MAPbI_3$ single-crystal planar devices, we fabricated control samples using (i) Au metal electrodes and (ii) graphite-based electrodes, and compared their current-voltage characteristics in the dark and under LED illumination. Specifically, the device with gold contacts were fabricated by transferring $MAPbI_3$ single crystals on prepatterned Cr/Au (5/60 nm) interdigitated metal electrodes (fabricated on $Si^{++}/SiO_2$ (280 µm/300 nm) substrates) using a Nitto ELP BT-150E-KL tape. Instead, the devices with graphite contacts were fabricated by directly depositing a custom made graphite conductive paste, realized by mixing 1.2 g of commercial graphite powder with 0,25 g of PMMA (polymethilmethacrylate), in 2,0 mL of chlorobenzene and leaving under stirring for 24 h.

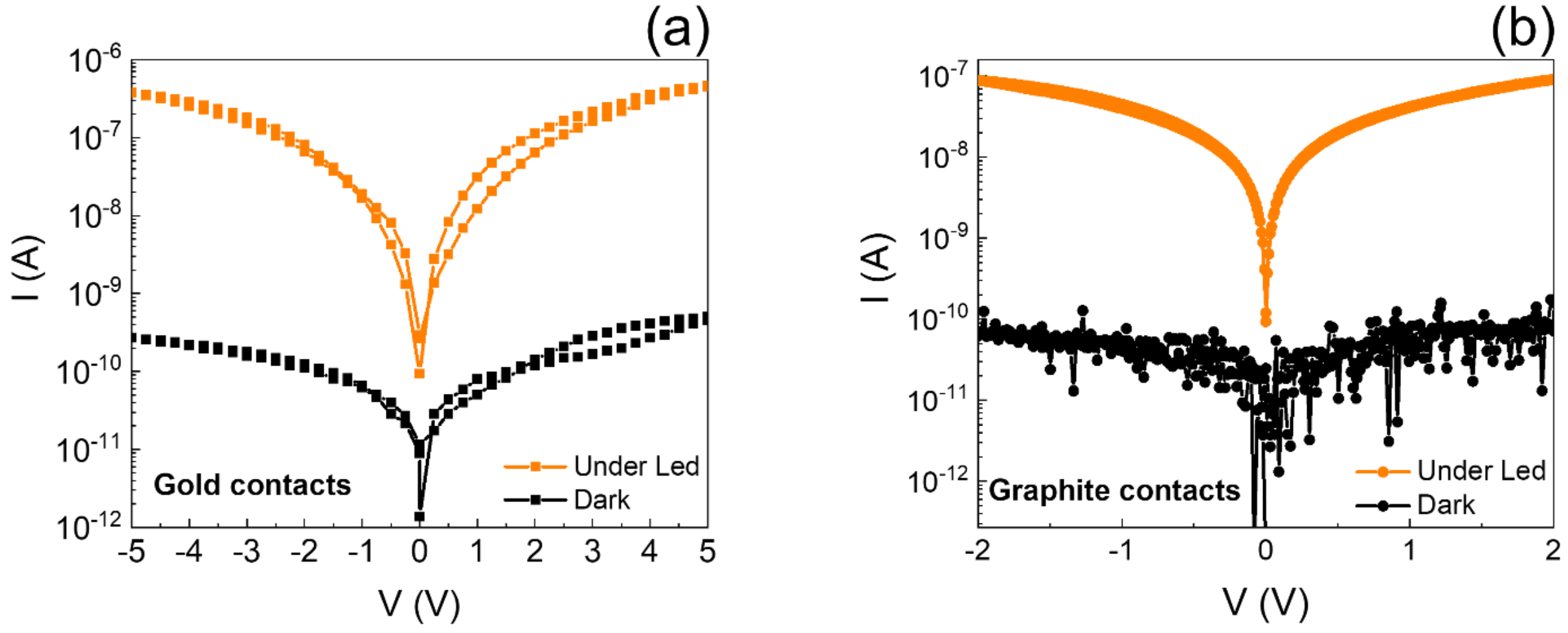

**Figure S2:** Semi-logarithmic I–V characteristics of planar $MaPbI_3$ single-crystal devices measured in the dark (black) and under LED illumination of 1.4 W (orange) for (a) Cr/Au contacts and (b) graphite contacts.

**Figure S2** shows representative semilogarithmic I-V curves acquired in the dark and under LED illumination for devices with Au (a) and graphite (b) contacts. In both cases, illumination increases the current by about 2-3 orders of magnitude in the explored bias range, indicating photo-enhanced conduction. Forward and reverse scans show no separation from the pronounced hysteretic switching activated by light observed in devices with Ag contacts. In particular, the I-V curves remain uniform and show no abrupt threshold-like transition under negative polarization in these control contacts.

## S.3 Back-to-back Schottky diode model for dark I-V

**Figure S3** reports in linear (a) and semi-log scale (b), the experimental data and fit curve (back-to-back Schottky diode model [1]) for the dark IV device characteristic at room temperature.

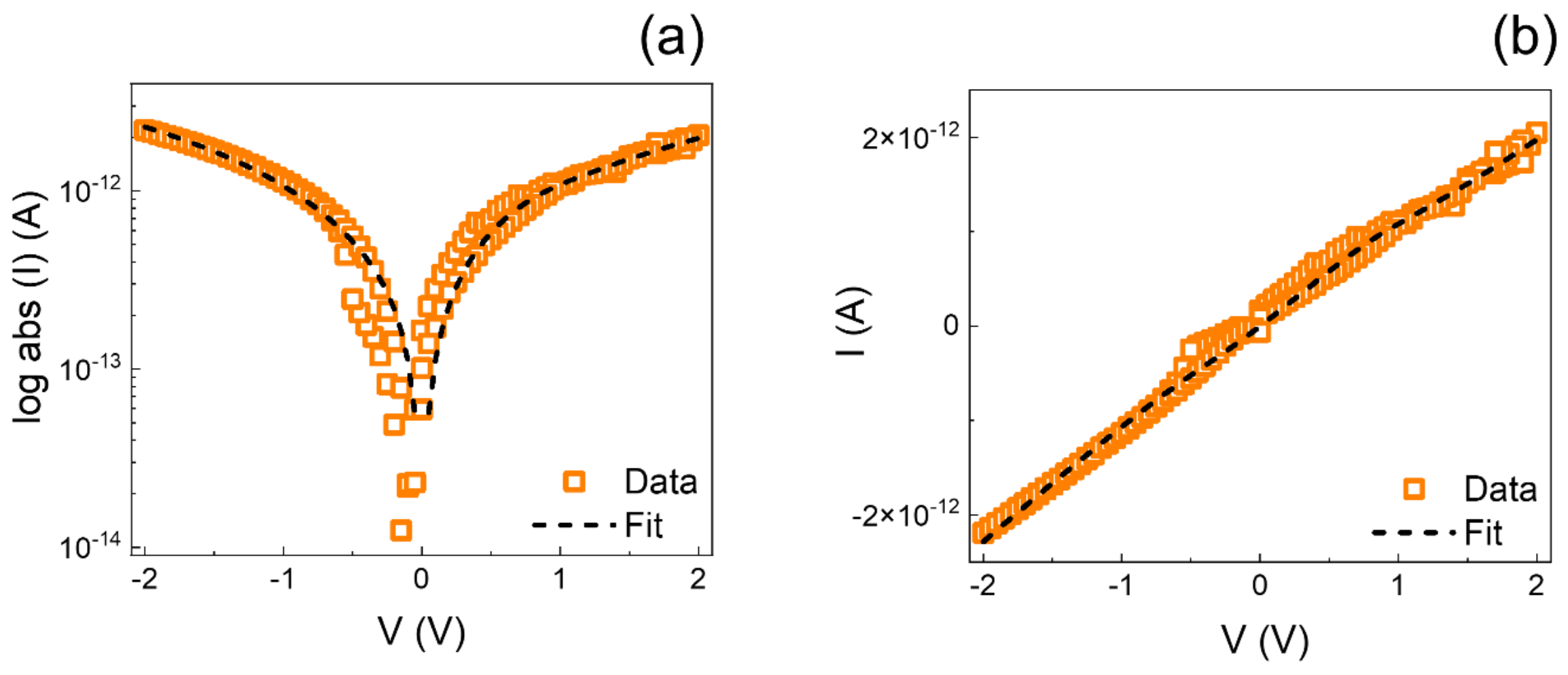

**Figure S3:** Dark I–V curve (orange) compared with the fit (red) for forward and (b) reverse branch obtained using a back-to-back Schottky diode model including series resistance $R_s$.

## S.4 Energy Dispersive X-ray Spectroscopy (EDS) analysis

Figure S4 a) reports the scanning electron micrograph of a pristine Ag/$MAPbI_3$ planar interface and b) the Ag line content profile in the direction perpendicular to the edge of the Ag electrode. In the pristine interface, the Ag signal is confined to the metal electrode region and rapidly drops to the background level in the perovskite region. Figure S4 c) and d) report the scanning electron micrograph of the Ag/MAPbI3 planar interface after electrical stress and transition to low resistance state and the Ag line content profile. In this case, the Ag signal remains slightly detectectable also in the perovskite region adjacent to the contact.

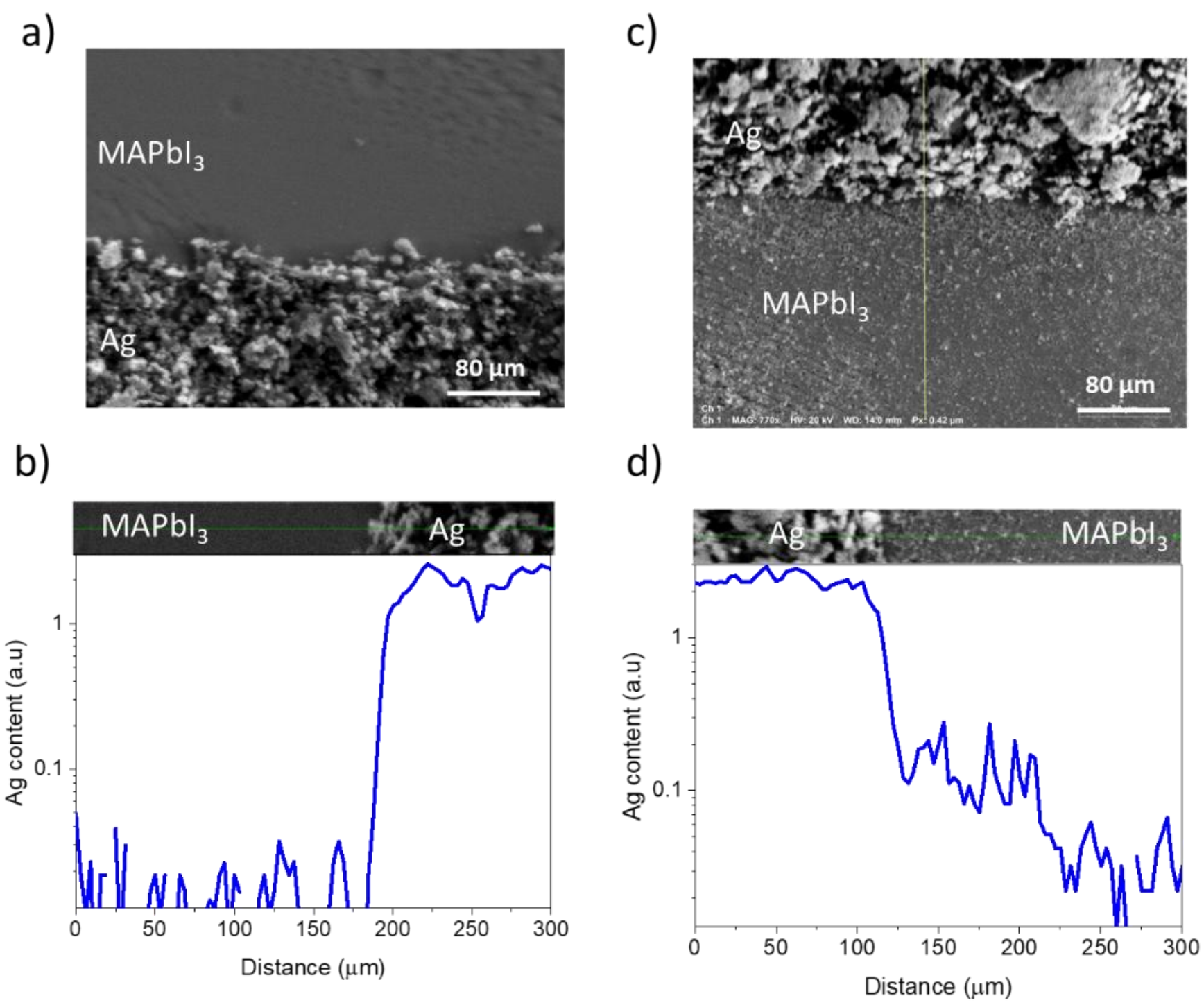

**Figure S4:** a) Scanning electron micrograph of a pristine Ag/$MAPbI_3$ planar interface and b) Ag line content profile in the direction perpendicular to the edge of the Ag electrode. c) Scanning electron micrograph of the Ag/$MAPbI_3$ planar interface after electrical stress and d) Ag line content profile in the direction perpendicular to the edge of the Ag electrode.

## S.5 Richardson-type law for negative-bias branch

To quantify the temperature dependence, the current was also extracted from negative-bias branch at $V = -0.5, -1$ and $-2\,V$ and analysed within a contact-limited thermionic-emission framework following a Richardson-type law [2,3] . **Figure S4** reports $ln(I/T^2)$ versus $1/T$ where the experimental data are well described by this linearization, allowing us to extract an effective Schottky barrier height $\Phi_B$. From the slopes, we obtain $\Phi_B = 0.6 \pm 0.03\ eV$ at $-0.5\ V$, $0.50 \pm 0.02\ eV$ at $-1\ V$, and $0.44 \pm 0.02\ eV$ at $-2\ V$ confirming the same overall picture of contact-limited thermally activated transport, extracted from positive bias, and a progressive reduction of the effective barrier with increasing bias magnitude.

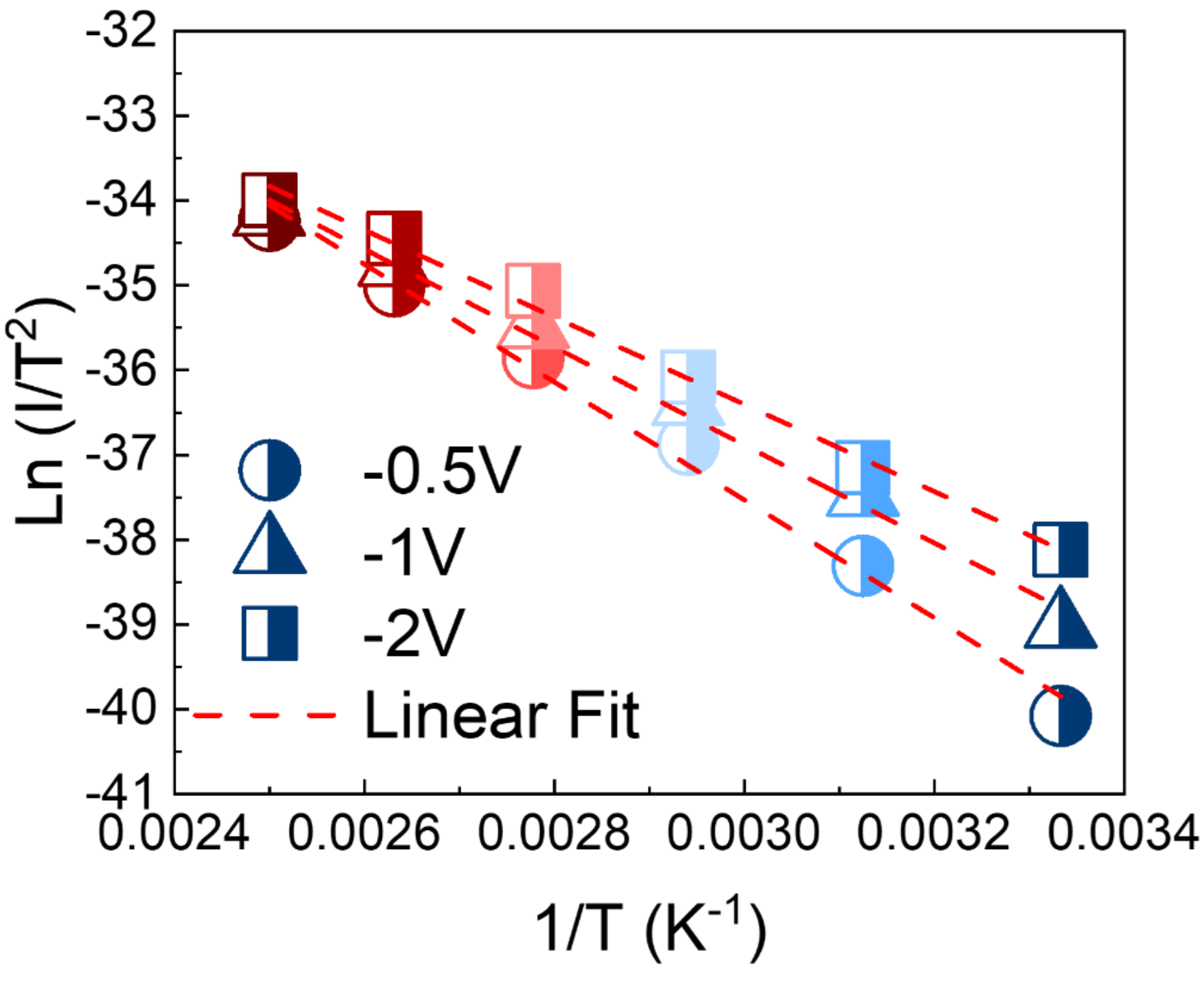

**Figure S4:** Richardson plot, $\ln(I/T^2)$ vs $1/T$ for different voltage bias (-0.5, -1, -2 V).

## S.6 Fitting parameters of the back-to-back Schottky model

| Fit parameter | Illuminated HRS | Illuminated LRS |
|---|---|---|
| Rs (Ω) | $1.4 \times 10^9$ | $4 \times 10^8$ |
| $\phi B_{01}$ (eV) | 1.07 | 0.8 |
| $\phi B_{02}$ (eV) | 0.8 | 0.81 |
| n1 | 1.36 | 1.04 |
| n2 | 1.04 | 1.03 |

**Table S6:** Main fitting parameter used to reproduce the experimental data reported in Figure 3 in the main text.

The transition from HRS to LRS is mainly captured by a decrease of the effective series resistance and by a reduction of one effective Schottky barrier, while the second barrier remains nearly unchanged. This supports a contact-mediated switching mechanism dominated by one Ag/$MAPbI_3$ interface. We stress that these parameters should be regarded as effective fitting parameters describing the device state under illumination, rather than fixed microscopic constants of the two interfaces.